\documentclass[final]{aipproc}

\layoutstyle{8x11double}

\def\be{\begin{equation}}
\def\ee{\end{equation}}
\def\gs{\mathrel{
   \rlap{\raise 0.511ex \hbox{$>$}}{\lower 0.511ex \hbox{$\sim$}}}}
\def\ls{\mathrel{
   \rlap{\raise 0.511ex \hbox{$<$}}{\lower 0.511ex \hbox{$\sim$}}}}

\newcommand{\ba}{\begin{array}{c}}
\newcommand{\baz}{\begin{array}{cc}}
\newcommand{\barrr}{\begin{array}{rrr}}
\newcommand{\bad}{\begin{array}{ccc}}
\newcommand{\bav}{\begin{array}{cccc}}
\newcommand{\baf}{\begin{array}{ccccc}}
\newcommand{\bea}{\begin{equation} \begin{array}{c}}
\newcommand{\eea}{ \end{array} \end{equation}}
\newcommand{\ea}{\end{array}}
\newcommand{\D}{\displaystyle}

\newcommand{\dm}{\mbox{$\Delta m^2$}}
\newcommand{\adm}{\mbox{$\overline{\Delta m^2}$}}
\newcommand{\meff}{\mbox{$\langle m \rangle$}}

\newcommand{\gsim}{\raise0.3ex\hbox{$\;>$\kern-0.75em\raise-1.1ex\hbox{
   $\sim\;$}}} 
\newcommand{\lsim}{\raise0.3ex\hbox{$\;<$\kern-0.75em\raise-1.1ex\hbox{
   $\sim\;$}}}

\begin{document}

\title{Neutrino Phenomenology of gauged $L_\mu - L_\tau$: 
\newline
MINOS and beyond}

\classification{ 14.60.St; 14.60.Pq; 14.70.Pw}
\keywords      {Neutrino; $Z'$; long range interaction}

\author{Julian Heeck}{}

\author{Werner Rodejohann}{
  address={Max--Planck--Institut f\"ur Kernphysik, Postfach 103980, 
D--69029 Heidelberg, Germany}
}

\begin{abstract}
If a $Z'$ gauge boson from a gauged $L_\mu - L_\tau$ symmetry is very
light, it is associated with a long-range leptonic force. In this case 
the particles in the Sun create via mixing of the $Z'$ with the
Standard Model $Z$ a flavor-dependent potential for 
muon neutrinos in terrestrial long-baseline experiments. The 
potential changes sign
for anti-neutrinos and hence can lead to apparent differences in
neutrino and anti-neutrino oscillations without introducing CP or CPT 
violation. This could for instance explain the recently found 
discrepancy in the MINOS experiment. 
We obtain the associated parameters of 
gauged $L_\mu - L_\tau$ required to explain this anomaly. 
The consequences for future long-baseline experiments are also
discussed, and we compare the scenario to standard NSIs. When used to explain 
MINOS, both approaches have severe difficulties with existing limits.

\end{abstract}

\maketitle


\section{Gauged $L_\alpha - L_\beta$ and Neutrinos}

In the Standard Model one can gauge one of the three lepton numbers 
$L_e - L_\mu$, $L_e - L_\tau$ or $L_\mu - L_\tau$ without introducing
anomalies \cite{Foot:1990mn}. The $U(1)$ gauge symmetry associated
with $L_\alpha - L_\beta$ goes along with a $Z'$ vector boson, which couples
to the current 
\begin{equation}
j'^\mu = \bar{\alpha} \, \gamma^\mu \, \alpha 
+ \bar{\nu}_\alpha \, \gamma^\mu
\, P_L \, \nu_\alpha - \bar{\beta} \, \gamma^\mu \, \beta 
- \bar{\nu}_\beta \, \gamma^\mu\, P_L \, \nu_\beta
\end{equation}
with coupling strength $g'$. Here $\alpha$ are the charged leptons and
$\nu_\alpha$ the corresponding neutrino. 
There is a priori no expectation for the mass of
the $Z'$. Here we will assume that the $Z'$ is ultra-light:  
$M_{Z'} < 1/R_{\rm A.U.} \simeq 10^{-18}$ eV, where 
$R_{\rm A.U.}$ denotes 
an astronomical unit.  In this case a Coulomb-like potential for
leptons, in particular neutrinos, is generated by the particles in the
Sun (and Earth). For instance, if we gauge $L_e - L_\beta$ 
one has   \cite{Joshipura:2003jh,Bandyopadhyay:2006uh} 
\begin{equation} \label{eq:Vem} 
V = \alpha_{e\beta} \, \frac{N_e}{R_{\rm A.U.}} \simeq 
1.3 \times 10^{-11} \, \left(\frac{\alpha_{e\beta}}{10^{-50}} \right)
{\rm eV}\,  , 
\end{equation}
where $\alpha_{e\beta} = g'^2/(4\pi)$, and $N_e$ is the number of
electrons in the Sun. In the 2-neutrino system of $\nu_e$ and
$\nu_\beta$ we have to add this potential to the usual oscillation
Hamiltonian:  
\[ 
{\cal H}_{e\beta} 
= \frac{\Delta m^2}{4 \, E} 
\left( \baz 
-\cos 2\theta & \sin 2\theta \\
\sin 2 \theta & \cos 2 \theta 
\ea \right) + 
\left( \baz 
V & 0 \\
0 & -V
\ea \right). 
\] 
The effect of this new neutrino physics looks very much like the
usually considered Non-Standard Interactions (NSIs), but does not depend
on the matter density and therefore would work even for vacuum
oscillations. The effect of $V$ on the mixing observables is 
{\setlength\arraycolsep{1.4pt}
\begin{eqnarray} 
\sin^2 2 \theta_V & 
= &  \D \frac{\sin^2 2 \theta}{1 - 4 \, \eta \, \cos 2
\theta + 4 \, \eta^2} \,  , \label{eq:Vmix} \\ 
 \Delta m_V^2 & 
= & \Delta m^2 \sqrt{1 - 4 \,  \eta \, \cos 2 \theta + 4 \,
\eta^2 } \label{eq:Vmas}\, ,
\end{eqnarray} 
}where $\eta = 2 \, E \, V/\Delta m^2$. 
Note that $V$ changes sign for anti-neutrinos, 
and hence an apparent difference between neutrino and anti-neutrino 
parameters will be measured. Note further that neither CP nor CPT
violation is required for this effect. From Eqs.~(\ref{eq:Vmix}, 
\ref{eq:Vmas}) it is seen that the mixing angle is required to be
non-maximal in order to introduce differences between neutrinos and 
anti-neutrinos. In the limit of small $\eta$ we have 
{\setlength\arraycolsep{1.4pt}
\begin{eqnarray}
\Delta m_V^2 - \overline{\Delta m_V^2}  & \simeq &  
- 4 \, \Delta m^2 \, \eta \, \cos 2 \theta \, , \\ 
\sin^2 2 \theta_V - \sin^2 2 \overline{\theta}_V
  & \simeq & 8 \, \eta \, 
\cos 2 \theta  \, \sin^2 2 \theta\, . 
\end{eqnarray} 
}In Eq.~(\ref{eq:Vem}) we have given the potential in units of very
small $\alpha_{e\beta}$. This is because the potential should be smaller than 
the energy scale $\Delta m^2/(4 \,E)$, which is about 
$6 \times 10^{-13} \left(\frac{\rm GeV}{E} \right)$ eV for
atmospheric neutrinos and 
$2 \times 10^{-11} \left(\frac{\rm MeV}{E}
\right)$ eV for solar neutrinos. With these estimates one can
understand the limits of  $\alpha_{e\mu} ~(\alpha_{e\tau})
\le 5.5~(6.4) \times 10^{-52}$ from atmospheric 
neutrinos \cite{Joshipura:2003jh}, and  $\alpha_{e\mu} 
~(\alpha_{e\tau}) \le 3.4~(2.5) \times
10^{-53}$ from solar and KamLAND neutrinos \cite{Bandyopadhyay:2006uh}. These
limits are more than one order of magnitude stronger than limits from
tests of the equivalence principle. 

We note here that in the symmetric limit the neutrino mass matrices for  
$L_e - L_\mu$ and $L_e - L_\tau$ conservation are  
\be 
m_\nu = \left( 
\bad 
0 & a & 0 \\
\cdot & 0 & 0 \\
\cdot & \cdot & b 
\ea \right) \mbox{ and } 
 \left( 
\bad 
0 & 0 & a \\
\cdot & b & 0 \\
\cdot & \cdot & 0 
\ea \right), 
\ee
respectively. Rather peculiar breaking patterns are
required to achieve successful neutrino mixing phenomenology from these
matrices. In contrast, if $L_\mu - L_\tau$ is conserved one has \cite{CR}
\be
m_\nu = 
\left( 
\bad 
a & 0 & 0 \\
\cdot & 0 & b \\
\cdot & \cdot & 0 
\ea \right). 
\ee
This matrix is automatically $\mu$--$\tau$ symmetric 
($\theta_{13} = |\theta_{23} - \pi/4| = 0$), hence requires 
less peculiar breaking,  and predicts the
presence of neutrino-less double beta decay ($\meff = a$). 
The masses are $a$ and $\pm b$, hence
neutrinos will have a mild, if any, hierarchy ($a \sim b$ because both
terms are allowed by the symmetry and therefore
expected to be of similar magnitude). 

\begin{figure}[t]
\centering
\includegraphics[height=.13\textheight,width=0.235\textwidth]{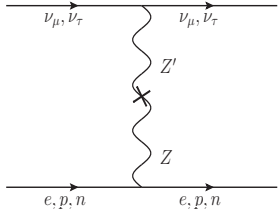}
\caption{Long-range $\nu_{\mu,\tau}$--$(e,p,n)$ interaction 
through $Z$--$Z'$-mixing.}
\label{fig:0}
\end{figure}

The question is now how to apply gauged  $L_\mu - L_\tau$ to neutrino
oscillations, because the lack of reasonable amounts of 
muons or tauons in the Universe
seems to forbid the generation of a potential in analogy 
to Eq.~(\ref{eq:Vem}). The solution \cite{HR} lies in $Z$--$Z'$
mixing, which in turn originates from 
the last two terms of the general Lagrangian 
{\setlength\arraycolsep{1.4pt}
\begin{eqnarray}
{\cal L} & = & -\frac{1}{4} {Z}'_{\mu\nu}\, 
{Z}'^{\mu\nu}+ \frac{1}{2} {M}_Z'^2 \, {Z}'_\mu \, {Z}'^\mu
- {g}' \, j'^\mu \, {Z}'_\mu \\  
& & -\frac{\sin \chi}{2} \, {Z}'^{\mu\nu}\, 
{B}_{\mu\nu} + \delta {M}^2 \, {Z}'_\mu \, {Z}^\mu \, . 
\end{eqnarray}
}Here ${Z}'_{\mu\nu}$ and ${B}_{\mu\nu}$ are the field strength tensors
of the new $U(1)$ and the Standard Model hypercharge. Diagonalizing the
kinetic and mass terms to obtain the physical particles $Z_{1,2}$ 
introduces $Z$--$Z'$ mixing: 
{\setlength\arraycolsep{1.4pt}
\begin{eqnarray}\nonumber 
 {\cal L}_{Z_1}  & = &   - \left(\frac{ e}{{s}_W {c}_W} 
{ 
\left( (j_3)_\mu -
{s}_W^2 \, (j_{\rm EM})_\mu \right)}  + g' \, \xi \, (j')_\mu\right) 
Z_1^\mu \, ,  \\ \nonumber 
	{\cal L}_{Z_2}  & = & 
- \left( g' \, (j')_\mu -  
\frac{e}{{s}_W \, {c}_W} 
\,  (\xi - s_W \, \chi)   \left( (j_3)_\mu 
 \right. \right.  \\ 
 & &  \left.  \left.  
 - {s}_W^2 \, (j_{\rm EM})_\mu
\right) - e \, c_W \, \chi \, 
(j_{\rm EM})_\mu \right) Z_2^\mu \, , \nonumber 
\end{eqnarray}
}where $\xi$ is a small mixing angle depending on $\chi$ and 
$\delta {M}^2$. The $Z'$ couples weakly with the electromagnetic and 
isospin currents $j_{\rm EM}$ and $j_3$, and mixes with the (mainly)
Standard Model $Z$. One can now obtain  \cite{HR} the following potential for 
$\nu_\mu$ and $\nu_\tau$ (see Fig.~\ref{fig:0}): 
\be \label{eq:V}
V = {\alpha} \, 
\frac{e}{4 \, s_W  c_W}
\frac{N_n}{4\pi \, R_{A.U.}} = 3.60\times 10^{-14} 
\, {\rm eV} \left(\frac{\alpha}{10^{-50}}\right)\, , 
\ee
where we have defined $\alpha = g' \, (\xi -s_W \, \chi) $ and
included the Earth's contribution to the solar one. 
For neutral objects like the Sun or Earth 
the electron and proton numbers cancel and only the neutron number
$N_n$ is of interest. The above
potential acts on the $\mu$--$\tau$ neutrino sector and introduces different
oscillation probabilities for neutrinos and anti-neutrinos. Consequently
it is a good candidate for an explanation of the MINOS results, which seemingly give
different mixing parameters in the muon neutrino and anti-neutrino 
survival probabilities. 

\begin{figure}[t]
\includegraphics[height=.18\textheight,width=0.235\textwidth]{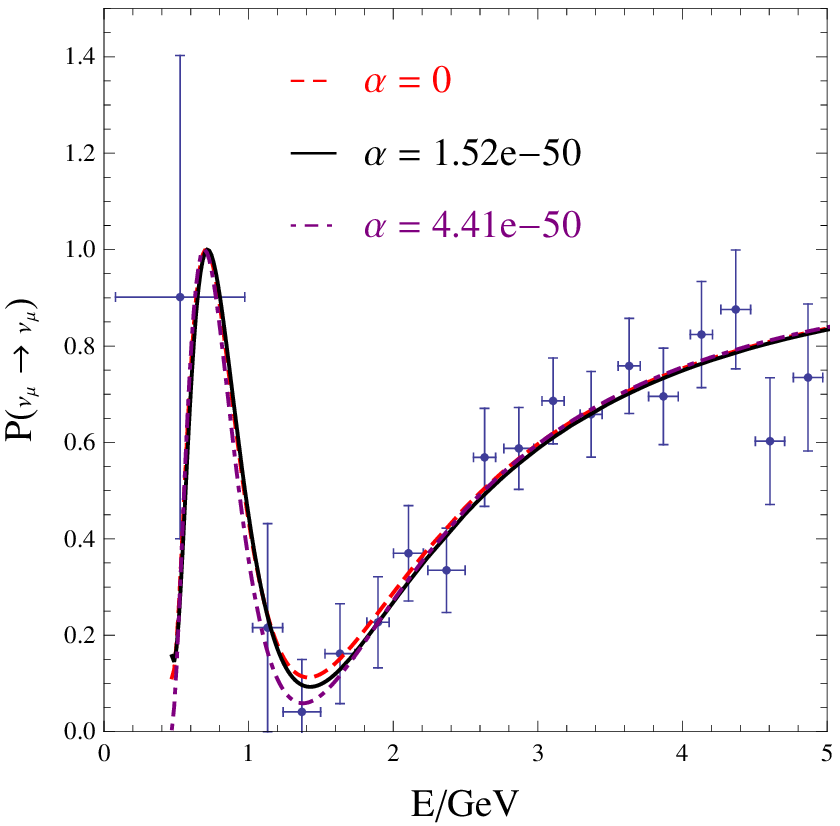}
\includegraphics[height=.18\textheight,width=0.235\textwidth]{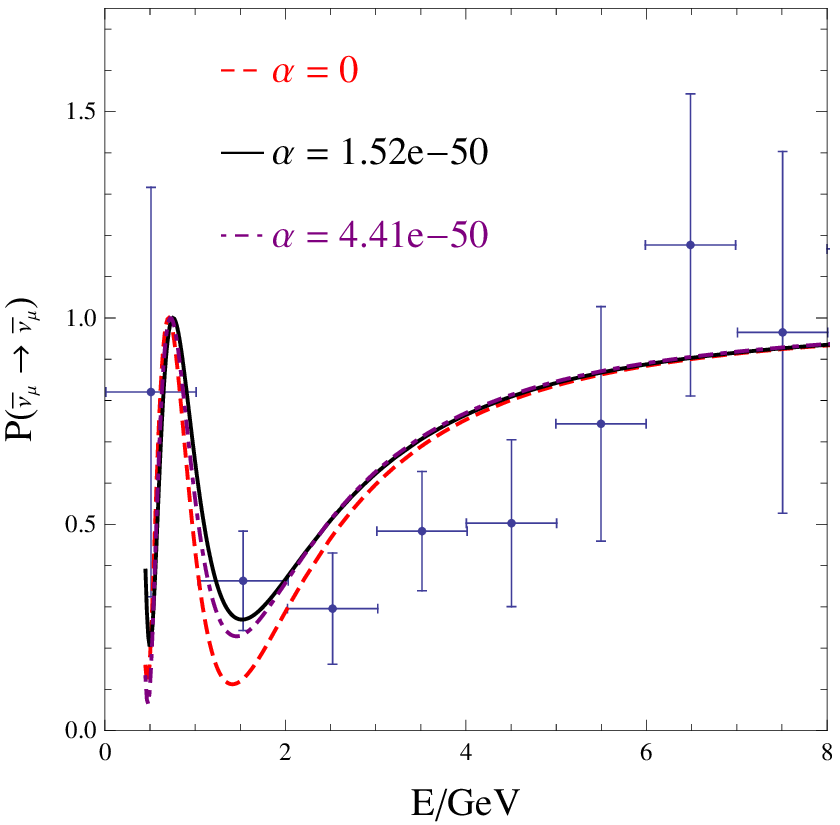}
  \caption{\label{fig:1}The oscillation probabilities for the best-fit values from
Eq.~(\ref{eq:fit}) for neutrinos and
anti-neutrinos superimposed on the MINOS data. Also plotted are  
the cases $\alpha = 0$ and the value for a second, local 
$\chi^2$-minimum. Taken from \cite{HR}.}
\end{figure}

\section{Application to MINOS and other Experiments}
The MINOS long-baseline experiment reported on individual measurements
of $\nu_\mu$ and $\bar{\nu}_\mu$ survival probabilities, and 
gave the following results \cite{minos}
\[
\ba 
\dm = \left(2.35^{+0.11}_{-0.08}\right) 
\times 10^{-3} \, {\rm eV}^2 ~,~~\sin^2 2
\theta > 0.91 \, , \\  \nonumber 
\adm = \left(3.36^{+0.45}_{-0.40}\right)  \times 10^{-3}\, {\rm eV}^2~,~~
 \sin^2 2 \overline{\theta} = 0.86 \pm 0.11 \, .  
\ea
\]
The apparent difference of the neutrino and anti-neutrino 
parameters has motivated several explanation attempts, in the form of 
CPT violation \cite{cpt}, NSIs \cite{NSI1,NSI2,ET}, sterile neutrinos 
plus gauged $B-L$ \cite{BL}, and gauged $L_\mu - L_\tau$ \cite{HR}. 
As became clear during this
meeting \cite{OY}, none of the explanations put forward so far 
works\footnote{An exception is probably CPT violation, if one is
willing to abandon such an important cornerstone of modern physics.}: 
the standard three-neutrino picture is remarkably stable
and robust. 

Let us illustrate the problems of the solutions: 
Fig.~\ref{fig:1} shows our fit to the MINOS data with the potential
from Eq.~(\ref{eq:V}). The best-fit values and $1\sigma$ ranges are \cite{HR}
\bea \label{eq:fit}
\sin^2 2 \theta = 0.83 \pm 0.08~,~~
\alpha = \left(1.52_{-1.14}^{+1.17}\right) \times 10^{-50} ~,\\
\Delta m^2 = (-2.48 \pm 0.19) \times 10^{-3}\,{\rm eV}^2  \, , 
\eea
with $\chi^2_{\rm min}/N_{\rm dof} = 47.77/50 \simeq 0.96$, to be
compared with the fit without new physics, which has $\chi^2_{\rm min}/N_{\rm
dof} = 49.43/51 \simeq 0.97$. Recall now that the total Hamiltonian
including $V$ looks like a typical NSI Hamiltonian, for which limits
have of course been derived already \cite{lim}. Values of 
$\alpha = 10^{-50}$ correspond to Earth matter NSIs of 
$|\epsilon_{\mu\mu}^\oplus| \simeq 0.25$. The current limit on this
parameter is $|\epsilon_{\mu\mu}^\oplus| \ls 0.068$, corresponding to 
$\alpha \ls 10^{-51}$, too small to have an effect of necessary size
for MINOS.  

However, there is one important difference to NSIs: in a gauge
invariant framework the $\epsilon$ parameters of the neutrino NSIs 
are responsible also for
charged lepton decays, which are subject to stringent constraints and
improve the bounds by typically one or two orders of magnitude. 
Of course, there might be an additional 
symmetry protecting the charged leptons, or highly fine-tuned 
cancellations of different higher order terms may take place \cite{tun}. 
For instance, consider the Lagrangian\footnote{This is a charged
current (CC) NSI, because neutral current NSIs required to explain the
MINOS data are at least of order 0.1 and hence in conflict with
bounds obtained from neutrino data alone \cite{NSI1,NSI2,ET}.} 
$ \mathcal{L}_{\rm CC}^{\rm NSI} \supset -2\sqrt{2} \, G_F \, 
\epsilon^d_{\tau\mu} \, V_{ud} \, 
  [\bar{u} \, \gamma^\mu \, d] \, [\bar{\mu} \, \gamma_\mu \, P_L \, \nu_\tau] $,
which leads to interference between $\nu_\mu$ CC events and events in
which $\nu_\mu$ oscillate into $\nu_\tau$, subsequently creating muons via
$\epsilon^d_{\tau\mu}$. 
For anti-neutrinos, $\epsilon^d_{\tau\mu} \to
(\epsilon^d_{\tau\mu})^\ast$, and hence different neutrino and
anti-neutrino parameters arise. Values of $|\epsilon^d_{\tau\mu}|$
around 0.1 are enough to explain the MINOS results. However, the Lagrangian
written in a gauge invariant way induces the tree-level decay $\tau \to
\mu \, \pi^0$, from which a limit of  $|\epsilon^d_{\tau\mu}| \ls
10^{-4}$ is derived \cite{BG}. We note here that the scenario of 
gauged $L_\mu - L_\tau$ discussed here does not suffer from such 
problems (the reason being diagonal and small couplings to leptons), 
and does not require strong and fine-tuned cancellations or
extra symmetries protecting charged leptons. 


Returning to neutrinos, a GLoBES \cite{globes} analysis of
future prospects for constraints on gauged $L_\mu - L_\tau$ has been
performed in \cite{HR}. Modifying
the program with the (now 3-flavor) Hamiltonian including $V$ from
Eq.~(\ref{eq:V}) and using the standard ``AEDL-files'' provided with
the software, we find future limits on $\alpha$ listed in Table
\ref{tab:1}.

\begin{table}[t]
\begin{tabular}{lr}
\hline
   \tablehead{1}{c}{b}{Experiment}
  & \tablehead{1}{c}{b}{Sensitivity to \\ $\alpha/10^{-50}$ at
$99.73\%$ C.L.} \\ \hline 
T2K ($\nu$-run) & $11.8$\\
T2K & $4.3$\\
T2HK & $1.7$\\
SPL & $7.5$\\
NO$\nu$A & $1.9$\\
Combined Superbeams & $1.4$\\
Nufact & $0.53$\\
\hline
\end{tabular}
\caption{Sensitivity to $\alpha$ from future experiments using GLoBES.}
\label{tab:1}
\end{table}

In Ref.~\cite{HR} a variety of experimental observables which could be
modified by the parameters of gauged $L_\mu - L_\tau$ is checked for
consistency. These include the magnetic moment of the muon, Big Bang
Nucleosynthesis, charge difference of electron and muon, electroweak
precision data, and tests of the equivalence principle. The strongest
constraints are and will be provided by neutrino oscillation
experiments, which shows the remarkable sensitivity of neutrinos to
new and interesting physics.


\begin{theacknowledgments}
WR thanks A.~Dighe, A.~Joshipura, T.~Schwetz and J.~Valle for discussions. 
This work was supported by the ERC under the Starting Grant 
MANITOP and by the DFG in the project RO 2516/4-1 as well as in the 
Transregio 27.

\end{theacknowledgments}

\bibliographystyle{aipprocl} 






\end{document}